\documentclass[aps,prd,amssymb,amsmath,nobibnotes,nofootinbib,superscriptaddress, reprint, 11pt]{revtex4}
\usepackage{amsfonts,amsmath,calrsfs, hyperref,url, color}
\allowdisplaybreaks
\usepackage{bm, bbm}
\usepackage{graphicx}
\usepackage{mathtools}
\usepackage{t1enc}
\usepackage{hepnames}
\usepackage{tensor}
\usepackage{comment}
\usepackage{appendix}

\usepackage{natbib}
\usepackage[dvipsnames]{xcolor}
\usepackage{tikz}
\usepackage{tikz-feynman}
\tikzfeynmanset{compat=1.0.0}

\allowdisplaybreaks

\usepackage{dcolumn}

\usepackage{adjustbox,booktabs}

\usepackage{makecell}
\usepackage{array}

\newcommand{\bc}{\begin{center}}
\newcommand{\ec}{\end{center}}
\newcommand{\be}{\begin{eqnarray}}
\newcommand{\ee}{\end{eqnarray}}
\newcommand{\bs}{\begin{slide}}
\newcommand{\es}{\end{slide}}

\newcommand{\bi}{\begin{itemize}}
\newcommand{\ei}{\end{itemize}}

\begin{document}
\title{Cutkosky rules and 1-loop $\kappa$-deformed amplitudes}

\author{Andrea Bevilacqua}
\affiliation{National Centre for Nuclear Research, ul. Pasteura 7, 02-093 Warsaw, Poland}

\date{\today}

\begin{abstract}

In this paper we show that the Cutkosky cutting rules are still valid term by term in the expansion in powers of $\kappa$ of the $\kappa$-deformed 1-loop correction to the propagator. We first present a general argument which relates each term in the expansion to a non-deformed amplitude containing additional propagators with mass $M>\kappa$. We then show the same thing more pragmatically, by reducing the singularity structure of the coefficients in the expansion of the $\kappa$-deformed amplitude, to the singularity structure of non-deformed loop amplitudes, by using algebraic and analytic identities. We will explicitly show this up to second order in $1/\kappa$, but the technique can be generalized to higher orders in $1/\kappa$. Both the abstract and the more direct approach easily generalize to different deformed theories. We will then compute the full imaginary part of the $\kappa$-deformed 1-loop correction to the propagator in a specific model, up to second order in the expansion in $1/\kappa$, highlighting the usefulness of the approach for the phenomenology of deformed models. This explicitly confirms previous qualitative arguments concerning the behaviour of the decay width of unstable particles in the considered model.

\end{abstract}

\maketitle

\section{Introduction}

The unification of quantum mechanics (QM) and general relativity (GR) has played a central role in theoretical physics for many decades (for a historical panoramic see \cite{Rovelli:2000aw}). Both theories are very successful within their respective domains, but they are based on very different physical assumptions, which in turn manifest themselves into their mathematical description. The unification of the two framework in a single theory of quantum gravity (QG), therefore, poses significant challenges both from a physical point of view, and from a more technical, mathematical point of view.

An interesting line of research originated from the pursuit of a final QG theory is represented by potential effective low-energy theories of a more fundamental QG model, such as those based on non-commutative spacetimes \cite{Lukierski:1992dt, Lukierski:1993wxa, Lukierski:1993wx, Kosinski:1994je, Kosinski:1994br, Lukierski:1996ib, Amelino-Camelia:2000stu, Amelino-Camelia:2002cqb, Amelino-Camelia:2002uql, Kowalski-Glikman:2001vvk}.  An advantage of this approach over more direct ones (in which QG models are built directly) is that it allows for an easier contact with phenomenology, for example in the context of astrophysical neutrinos \cite{Amelino-Camelia:2016fuh, Amelino-Camelia:2016wpo, Amelino-Camelia:2016ohi, Rosati:2017hod, Amelino-Camelia:2022pja}. In this paper we will focus on the $\kappa$-Minkowsky spacetime, with commutation relations $[x^0, \mathbf{x}] = i\frac{\mathbf{x}}{\kappa}$ (for a review, see for example \cite{Kowalski-Glikman:2017ifs, Majid:1994cy, Arzano:2021scz} and references therein). In this model, the quantity $\kappa$ has the dimension of mass, and it is usually identified with the Planck mass $m_p$. As such, it is often sufficient to calculate the first order correction in the small parameter (in our case proportional to $1/\kappa$) to canonical, non-deformed quantities to obtain predictions of such deformed models. 

In previous publications \cite{Bevilacqua:2023pqz, Arzano:2020jro, Bevilacqua:2022fbz}, we developed and analyzed a model of a (complex) scalar field in \(\kappa\)-Minkowski spacetime, with a particular emphasis on its discrete symmetries — charge conjugation (C), parity (P), and time reversal (T) — as well as the algebra of the associated Noether charges. One interesting aspect of this model is obtained in the study of the decay times of particles and antiparticles, in particular in different reference frames \cite{Bevilacqua:2022fbz, Bevilacqua:2023swc, Bevilacqua:2024jpy}. Due to the nature of deformation, even non-interacting models can exhibit non-trivial phenomenology. A key point in the investigation of the decay of particles and antiparticles is the determination of higher order corrections to the propagator of the decaying particle (and in particular their imaginary part). Some preliminary results in this sense have been obtained in \cite{Bevilacqua:2023tre}, where (for simplicity) only the imaginary part coming from the logarithms was calculated.

In particular, in this paper we focus on the Cutkosky cutting rules for Feynman diagrams in order to compute the imaginary part of the loop correction. In non-deformed QFT, they provide a pragmatic way of obtaining the imaginary part of Feynman amplitudes (especially at 1-loop). In the literature, the validity of the Cutkosky rules has already been investigated for deformed models in $2+1$ dimensions \cite{Sasai:2009jm}, where it was possible to integrate the full deformed amplitude explicitly over momentum space. More general models (like the one considered here) lack this advantage. For the sake of definiteness, we will consider the model presented in \cite{Bevilacqua:2023pqz, Arzano:2020jro, Bevilacqua:2022fbz} and obtain the $\kappa$-deformed amplitude for the 1-loop correction to the propagator, expanded in powers of $1/\kappa$. One can then apply the Cutkosky rule term by term in the series. We first present a general way to relate the singularity structure of each term in the series, to the structure of a non-deformed diagram with additional heavy particles. To actually proceed with the computations, however, instead of applying the Cutkosky cutting rules directly, which may be cumbersome in the deformed context since they involve derivatives of Dirac deltas, we first use some analytic and algebraic identities to rewrite the amplitude in terms of simpler ones. We will obtain some `reduction identities', which relate the singularity structure of each term in the series in powers of $1/\kappa$ to non-deformed amplitudes, but in a different way than what has been discussed above. These will then allow us to use directly the Cutkosky rules, which can therefore be used term to term in perturbation, and shows their validity in the expanded amplitude. 

Our procedure is completely general. It can be applied to any theory with deformed sum of momenta which reduce to the canonical sum in the non-deformed limit, provided that the only singularities come from a deformation of the momentum conservation in the propagator. Applying the procedure to the model introduced in \cite{Bevilacqua:2023pqz, Arzano:2020jro, Bevilacqua:2022fbz}, after expanding the amplitude up to second order in powers of $1/\kappa$, allows us to obtain phenomenological predictions. This will allow us to explicitly obtain the first non-trivial correction due to deformation of the decay width for unstable particles, confirming the qualitative results obtained in \cite{Bevilacqua:2022fbz}. In particular, we show that the deformation correction to the decay width $\Gamma$ cannot be larger than $p_0^2/\kappa^2$ in this model ($p_0$ is the energy of the decaying particle), so that it can be neglected when extracting phenomenological consequences of deformation, see for example \cite{Bevilacqua:2024jpy, Bevilacqua:2022fbz}.

 {\section{Structure of the paper}}

 {The paper is structured around three main aspects. The first one presents the main theoretical results of the paper concerning the validity of the Cutkosky cutting rules in the deformed context. The second aspect shows some computations which serve either as preparation or as checks of the discussion or of the formulae presented along the first line. The third one is more phenomenology-oriented, and consists of an application of the discussion in the first line, leading to new physical results concerning the imaginary part of the $\kappa$-deformed 1-loop correction to the propagator. These aspects are intertwined in the paper, but they can be traced to the following sections:
\begin{align*}
    \text{aspect 1:} & \,\, \ref{defeypro} \rightarrow \,\, \ref{defamp} \rightarrow \,\, \ref{cutdef-dim} \rightarrow \,\, \ref{cutdef} \rightarrow \,\, \ref{conclusion}; \\
    \text{aspect 2:} & \,\, \ref{review} \rightarrow \,\, \ref{nondefloopcorr} \rightarrow \,\, \ref{checkdefcut}; \\
    \text{aspect 3:} & \,\, \ref{imcomp} \rightarrow \,\, \ref{conclusion}.
\end{align*}}
 {We now describe in more detail the content of each section of the paper. In section \ref{review} we quickly review the momentum-space picture of the $\kappa$-Minkowski spacetime.} In section \ref{defeypro} we will briefly discuss the Feynman propagator in the deformed context. In section \ref{non-def-im} we will then quickly review the computation of the imaginary part of the 1-loop correction to a scalar propagator (in a $\varphi \phi^2$ theory), both using the canonical approach, and using the Cutkosky cutting rules. Then, in section \ref{defamp} we will introduce our deformed amplitude for the 1-loop correction to the propagator, and we will expand up to second order in powers of $1/\kappa$. The Cutkosky rules are then discussed in section \ref{cutdef-dim}. In this section, we relate the singularity structure of each term in the expansion of the deformed amplitude, to the singularity structure of a non-deformed diagram, containing propagators with $M>\kappa$. We will then proceed to the more pragmatic approach in section \ref{cutdef}. In particular, we will obtain two reduction identities relating the expansion of the deformed amplitude in powers of $1/\kappa$, to non-deformed amplitudes. These relations are well defined provided that the integrals are finite, which is the case since we are interested in the imaginary part of the correction. Before moving to the actual computation of the imaginary part of the amplitude in section \ref{defamp}, in section \ref{checkdefcut} we will first provide an example of check of the validity of one of our reduction relations obtained in section \ref{cutdef}. We will then compute the imaginary part of the $\kappa$-deformed 1-loop correction to the amplitude in section \ref{imcomp}.  {Although the results of section \ref{defamp} hold in an arbitrary frame, for simplicity we will only} carry out the explicit computations in the centre of mass (c.o.m.) frame in section \ref{comframe}, and we will then argue about the result in a general frame in section \ref{generalframe}.  {This pragmatic approach will allow us to showcase the power of the method introduced in the paper, while still producing interesting results concerning the imaginary part of the 1-loop $\kappa$-deformed correction to the scalar propagator, both from a theoretical and phenomenological point of view.} We will then discuss and summarize our results in section \ref{conclusion}.

 { \section{Brief review of $\kappa$-deformation and momentum-space picture}\label{review} }
 {In this section we will briefly review the momentum-space picture of $\kappa$-Minkowski spacetime, referring the reader for example to \cite{Kowalski-Glikman:2017ifs, Majid:1994cy, Arzano:2021scz} for more additional details. The starting point is represented by an explicit representation of the Lie algebra $[x^0, \mathbf{x}] = i\frac{\mathbf{x}}{\kappa}$ (called $\mathfrak{an}(3)$ algebra) given by
\begin{align}\label{II.1.18}
	\hat x^0 = -\frac{i}{\kappa} \,\left(\begin{array}{ccc}
		0 & \mathbf{0} & 1 \\
		\mathbf{0}^T & \tilde{\mathbf{0}} & \mathbf{0}^T \\
		1 & \mathbf{0} & 0
	\end{array}\right) \quad
	\hat{x}^i = \frac{i}{\kappa} \,\left(\begin{array}{ccc}
			0 & {(\epsilon^i)\,{}^T} &  0\\
			\epsilon^i & \tilde{\mathbf{0}} & \epsilon^i \\
			0 & -(\epsilon^i)\,{}^T & 0
		\end{array}\right).
\end{align}
One can then build the plane waves (which are therefore also group elements of the $AN(3)$ group)
\begin{align}\label{defek}
	\hat{e}_k =e^{ik_i \hat x^i} e^{ik_0 \hat x^0}\,.
\end{align}
In the above expression, the $k_\mu$ are dimensionful quantities which can be interpreted as momentum-space coordinates, and are called bicrossproduct basis. In this representation, spacetime coordinates are matrices, and the product of plane waves is the group product of $AN(3)$. It is in general useful to switch to a more familiar representation of spacetime, in which coordinates are commutative and all deformation effects are translated to a non-trivial $\star$ product between plane waves, which in turn implies a non-trivial composition rule $\oplus$ and non-trivial antipode $S(\cdot)$ for momenta. This is done through a Weyl map $\mathcal{W}$ defined by
\begin{align}
	\mathcal{W}(\hat{e}_k  (\hat{x})) = e_{p  (k)}(x), 
	\qquad
	\mathcal{W}^{-1}(e_{p  (k)}(x)) = \hat{e}  _k(\hat{x}),
    \qquad
	e_{p  (k)(x)} = e^{-i(\omega_\mathbf{p}  t - \mathbf{p}  \mathbf{x})},
\end{align}
with the property that 
\begin{align}
	\mathcal{W}(({\hat{e}  }_k)^{-1}) = e_{S(p  (k))}, 
	\qquad
	\mathcal{W}(\hat{e}  _k\hat{e}  _l)
	=
	e_{p  (k)\oplus q  (l)} =: e_{p  } \star e_{q  }.
\end{align}
In other words, space and time non-commutativity is translated to a non-trivial momentum composition in canonical spacetime. The basis $p(k)$ used in $e_p$ is called classical basis. A full characterization of Weyl maps would lead us astray, a more in depth discussion of the Weyl maps and their subtleties can be found in \cite{Kowalski-Glikman:2017ifs, Arzano:2021scz, Arzano:2020jro}. In what follows we will use the classical basis for simplicity, but the discussion of the validity of the Cutkosky rules in the deformed context is completely general, since the only requirement is a non-trivial momentum composition rule which is a general feature of non-commutative spacetimes. 
}

\section{$\kappa$-deformed Feynman propagator}\label{defeypro}

The model into consideration is the $\kappa$-deformed complex scalar field introduced in \cite{Arzano:2020jro}. In this model, the action is given by 
\begin{align}
    S_{\text{free}}
	&=
	\frac{1}{2}
	\int_{\mathbb{R}^4} d^4x \,\, 
	(\partial^\mu \phi)^\dag \star (\partial_\mu \phi) 
	+
	(\partial_\mu \phi) \star (\partial^\mu \phi)^\dag  
	- 
	m^2 \phi^\dag \star \phi
	- 
	m^2 \phi \star \phi^\dag
\end{align}
and the field (and its Hermitian conjugate) satisfies the Klein-Gordon equation. 

In non-deformed QFT one can get the propagator in several equivalent ways, for example by directly computing the VEV of the time ordered product of two fields, as the Green function of the field operator, or by taking functional derivatives of the path integral. In the deformed context, the first of these approaches would require us to to compute objects like $\phi^\dag(x) \star \phi(y)$, which in turn implies that we need to make some assumption for $[x^0,\mathbf{y}^i]$. The second approach provides a straightforward computation, since following \cite{Amelino-Camelia:2001rtw}, \cite{Arzano:2018gii} one can use the definitions
\begin{align}
	Z[J, J^\dagger] 
	=
	\frac{1}{Z[0,0]}
	\int\mathcal{D}[\phi] \mathcal[\phi^\dag]
	e^{i S_{\text{free}}[\phi, \phi^\dag] 
		+ 
		\frac{i}{2}\int
		[
		\phi^\dag \star J + J \star \phi^\dag + J^\dag \star \phi + \phi \star J^\dag
		]
	}
\end{align} 
and the off-shell momentum-space action
\begin{align}
    S =
	\frac{1}{2}
	\int \frac{d^4p}{p_4^2/\kappa^2}\, \theta(p_0)\,
	\zeta^2(p)
	\left(
	1 + \frac{|p_+|^3}{\kappa^3}
	\right)
	(p_\mu p^\mu - m^2)
	[
	a^\dag_\mathbf{p} a_\mathbf{p}
	+
	b^\dag_{\mathbf{p}^*} b_{\mathbf{p}^*}
	]
\end{align}
to show that the propagator in momentum space is the same as the non-deformed one \cite{Bevilacqua:2023tre}. However, as shown in \cite{Bevilacqua:2023tre}, the above off-shell action can only be defined in some subset of momentum space, since one needs to ensure that off-shell one has $\omega_p \oplus \omega_q > 0$ and $S(\omega_p)\oplus S(\omega_q) <0$ (these inequalities are satisfied on-shell). For these reasons, we will postpone the discussion of these two approaches to future publications, and we concentrate on the propagator as the Green function of the field operator. Since the field operator is not deformed, we simply have that the propagator in momentum space is given by the canonical scalar propagator.

\section{Example of loop corrections to the propagator: non-deformed case}\label{nondefloopcorr}

\subsection{Quick review of the direct computation} \label{non-def-im}

We consider the following diagram in a $\psi\phi^2$ theory, where both $\psi$ and $\phi$ are scalars of mass $m_\psi$ and $m$ respectively.
\begin{figure}[h]
    \centering
    \includegraphics[width=0.3\linewidth]{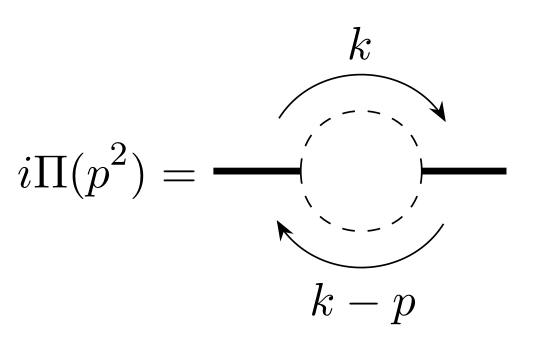}
    \label{fig:1}
\end{figure}
With interaction vertex $g\psi\frac{\phi^2}{2!}$, the vertex is given by $i g$ and the propagator is the canonical one. The amplitude can be written as follows.
\begin{align}\label{initialamp}
	i\Pi(p^2)
	=
	(ig)^2
	\int \frac{d^4 k}{(2\pi)^4} 
	\frac{i}{k^2 - m^2 +i\epsilon}
	\frac{i}{(p-k)^2 - m^2 +i\epsilon}.
\end{align}
Using Feynman parameters, performing the Wick rotation, and using the identity 
\begin{align}\label{dimregint}
	\mu^{4-d}
	\int
	\frac{d^d k}{(2\pi)^4}
	\frac{(k^2)^a}{(k^2 + \Delta)^b}
	=
	\frac{\Gamma(b-a-\frac{1}{2}d)\Gamma(a+\frac{1}{2}d)}{(4\pi)^{d/2}\Gamma(b)\Gamma(\frac{1}{2}d)}
	\Delta^{-(b-a-d/2)}\mu^{4-d},
\end{align}
the momentum space integral becomes
\begin{align}\label{epsilon2}
	\frac{\Gamma(\frac{\epsilon}{2})}{(4\pi)^2 \Gamma(2)}
	\Delta^{-\frac{\epsilon}{2}}
	(\mu^2)^{\frac{\epsilon}{2}}.
\end{align}
Upon expanding, it reduces to 
\begin{align}\label{undefresbeforeren}
	\Pi(p^2)
	=
	-\frac{(ig)^2}{(4\pi)^2}
	\int_0^1 dx \,
	\left[
	\frac{2}{\epsilon} - \log \frac{\Delta}{\mu^2} -\gamma + O(\epsilon)
	\right].
\end{align}
It is then sufficient to notice that the argument of the logarithm is positive unless $p^2 x(1-x)> m^2$, in which case the numerator becomes negative, and therefore the whole argument of the logarithm becomes negative. 
Therefore, as long as $L_1 < x<L_2$ where $L_1 < \frac{1}{2} - \frac{1}{2} \sqrt{1-4\frac{m^2}{p^2}}$ and $L_2 < \frac{1}{2} + \frac{1}{2} \sqrt{1-4\frac{m^2}{p^2}}$, and if $p^2>4m^2$, the amplitude gains a small imaginary part. In this case $\Im \log \frac{\Delta}{|\Delta^0|} = -\pi$, and we finally have
\begin{align}\label{undefim}
	\Im \Pi(p^2) = \frac{g^2}{16\pi}\sqrt{1-4\frac{m^2}{p^2}}.
\end{align}

\subsection{Quick review of the Cutkosky rules}

The Cutkosky rules \cite{Cutkosky:1960sp} allow one to compute the imaginary part of a diagram by summing over all the possible cuts of the diagram, were for each cut propagator one performs the substitution 
\begin{align}
	\frac{1}{D} \mapsto 2\pi \theta(p_0) i \delta(D).
\end{align}
Here we apply these rules to compute once again the imaginary part of the non-deformed loop. From our initial definition
\begin{align}
	i\Pi(p^2)
	=
	(ig)^2
	\int \frac{d^4 k}{(2\pi)^4} 
	\underbrace{\frac{i}{k^2 - m^2 +i\epsilon}}_{B}
	\underbrace{\frac{i}{(p-k)^2 - m^2 +i\epsilon}}_{A}
\end{align}
there is only one possible cut, the one which cuts both propagators of the loop. We get
\begin{align}
	2\Im [i\Pi(p^2)]
    =
	(ig)^2
	\int \frac{d^4 k}{(2\pi)^4} 
	d^4 q \,\,
	(2\pi i)^2
	\theta(k_0)\theta(q_0)
	\delta(k^2 - m^2)
	\delta(q^2 - m^2)
	\delta^4(q - (p-k))\label{prelim}
\end{align}
which can be rewritten as
\begin{align}
	2\Im [i\Pi(p^2)]
	=
	g^2
	\int \frac{d^3 k}{(2\pi)^2} 
	d^3 q \,\,
	\frac{1}{4k_0 q_0}
	\delta^4(q - (p-k))
	=
	\frac{g^2}{16\pi^2}
	\int d^3 k 
	\frac{1}{k_0 q_0}
	\delta(q_0 + k_0  - p_0).
\end{align}
It is convenient to switch to the center of mass frame, in which $p_\mu = (p_0, 0,0,0)$, and $k_\mu = (\sqrt{|\mathbf{k}|^2 + m^2}, \mathbf{k})$ and $q_\mu = (\sqrt{|\mathbf{k}|^2 + m^2}, -\mathbf{k})$ (recall that all the particles are on-shell). The energies of the two particles after the vertex are the same, and their sum is equal to the initial energy. This means that the above integral becomes (we also pass to polar coordinates)
\begin{align}
	2\Im [i\Pi(p^2)]
	&=
	\frac{g^2}{16\pi^2}
	\int d|\mathbf{k}|\,4\pi|\mathbf{k}|^2\, 
	\frac{1}{|\mathbf{k}|^2+ m^2}
	\delta(2\sqrt{|\mathbf{k}|^2+ m^2}  - p_0) \\
    &=
	\frac{g^2}{8\pi}
	\sqrt{1-4\frac{m^2}{p^2}}
\end{align}
where we used the fact that $p_0^2 = p^2$ in the center of mass frame. Therefore we get
\begin{align}\label{rescut1}
	\Im [i\Pi(p^2)]
	=
	\frac{g^2}{16\pi}
	\sqrt{1-4\frac{m^2}{p^2}},
\end{align}
which is of course the same result as before.

\section{The $\kappa$-deformed amplitude}\label{defamp}

To compute the 1-loop correction in the $\kappa$-deformed context, we assume that the vertex contribution remains $ig$, but the canonical momentum conservation is switched with the deformed one.  {More precisely, in analogy with the non-deformed case, we consider an interaction vertex containing the fields $\phi$ and $\varphi$. Because of the non-commutativity of the star product, we have three possibilities, namely
\begin{align}
    \phi \star \varphi\star \varphi,
    \qquad
    \varphi \star \phi \star \varphi,
    \qquad
    \varphi \star \varphi \star \phi,
\end{align}
corresponding to the following three possible momentum conservation relations (we consider the momentum of the $\phi$ particle as incoming for definiteness, ad we call $p_2$ and $p_3$ the momenta of the $\varphi$ particles)
\begin{align}
    S(p) \oplus p_2 \oplus p_3 = 0,
    \qquad
    p_2 \oplus S(p) \oplus p_3 = 0,
    \qquad
    p_2 \oplus p_3 \oplus S(p) = 0.
\end{align}
We now discuss all three possibilities.
The first conservation of momentum is immediately satisfied if $p_2 \oplus p_3 = p$,} and we can use one of the following parametrizations for $p_2$ and $p_3$
\begin{align}\label{p1}
    \left\{
	p_2 = k,
	\,\,
	p_3 = S(k)\oplus p
    \right\},
\quad
    \left\{
	p_2 = p\oplus S(k),
	\,\,
	p_3 = k
    \right\}
\end{align}
\begin{align}\label{p3}
    \left\{
	p_2 = S(k),
	\,\,
	p_3 = k \oplus p
    \right\},
\quad
    \left\{
	p_2 = p\oplus k,
	\,\,
	p_3 = S(k)
    \right\}
\end{align}
The conservation of momenta $p_2 \oplus p_3 \oplus S(p) = 0$ gives the same four possibilities. Furthermore, the additional possibility $p_2 \oplus S(p) \oplus p_3 = 0$ reduces to the previous two possibilities if we use braiding \cite{Arzano:2022vmh}. We will assume that this is the case, and we will not investigate this issue further in this paper. 

Therefore, the equations \eqref{p1}, \eqref{p3}  are all we need to take into consideration. We have therefore
\begin{align}\label{Pi1}
	i\Pi_1(p^2)
	=
	(ig)^2
	\int \frac{d^4 k}{(2\pi)^4} 
	\frac{i}{k^2 - m^2 +i\epsilon}
	\frac{i}{(S(k)\oplus p)^2 - m^2 +i\epsilon}
\end{align}
\begin{align}\label{Pi2}
	i\Pi_2(p^2)
	=
	(ig)^2
	\int \frac{d^4 k}{(2\pi)^4} 
	\frac{i}{k^2 - m^2 +i\epsilon}
	\frac{i}{(p \oplus S(k))^2 - m^2 +i\epsilon}
\end{align}
\begin{align}\label{Pi3}
	i\Pi_3(p^2)
	=
	(ig)^2
	\int \frac{d^4 k}{(2\pi)^4} 
	\frac{i}{k^2 - m^2 +i\epsilon}
	\frac{i}{(k \oplus p)^2 - m^2 +i\epsilon}
\end{align}
\begin{align}\label{Pi4}
	i\Pi_4(p^2)
	=
	(ig)^2
	\int \frac{d^4 k}{(2\pi)^4} 
	\frac{i}{k^2 - m^2 +i\epsilon}
	\frac{i}{(p\oplus k)^2 - m^2 +i\epsilon}
\end{align}
We now need to expand them. We proceed as follows (we show the procedure for the first propagator, and the others are performed in the same way)
\begin{align}
	\frac{i}{(S(k)\oplus p)^2 - m^2 +i\epsilon}
	&\approx
	\frac{i}{(S(k)\oplus p)^2 - m^2 +i\epsilon}
	\Big|_{\frac{1}{\kappa}= 0}
	+
	\frac{1}{\kappa}
	\frac{\partial}{\partial \frac{1}{\kappa}}
	\frac{i}{(S(k)\oplus p)^2 - m^2 +i\epsilon}
	\Big|_{\frac{1}{\kappa}= 0} \nonumber \\
	&+
	\frac{1}{2 \kappa^2}
	\frac{\partial^2}{\partial \left(\frac{1}{\kappa}\right)^2}
	\frac{i}{(S(k)\oplus p)^2 - m^2 +i\epsilon}
	\Big|_{\frac{1}{\kappa}= 0} 
\end{align}
Taking as example the first propagator, we have therefore
\begin{align}
	\frac{i}{(S(k)\oplus p)^2 - m^2 +i\epsilon}
	&=
	\frac{i}{(p-k)^2 - m^2 +i\epsilon}
	+
	\frac{1}{\kappa}
	\frac{i\Delta_1[(S(k)\oplus p)^2]}{[(p-k)^2 - m^2 +i\epsilon]^2} \nonumber \\
	&+
	\frac{1}{\kappa^2}
	\left[
	\frac{i(\Delta_1[(S(k)\oplus p)^2])^2}{[(p-k)^2 - m^2 +i\epsilon]^3}
	+
	\frac{i\Delta_2[(S(k)\oplus p)^2]}{[(p-k)^2 - m^2 +i\epsilon]^2}
	\right]
\end{align}
Notice that the additional minus signs is already included in the $\Delta_1$s, and the factor $1/2$ coming from the second derivative is already included in the $\Delta_2$s together with the additional minus sign.
The total amplitude is therefore given by
\begin{align}
	i\Pi_{TOT}(p^2)
	=
	\frac{i\Pi_1(p^2)+i\Pi_2(p^2)+i\Pi_3(p^2)+i\Pi_4(p^2)}{4}
\end{align}
where the factor $1/4$ is needed to get the correct $\kappa\rightarrow \infty$ limit, and we get
\begin{align}\label{defamplitude}
	i\Pi_{TOT}(p^2)
	&=
	i\Pi_U(p^2) \\
	&+
	\frac{g^2}{\kappa}
	\int \frac{d^4k}{(2\pi)^4}
	f_1(k)
	\frac{1}{k^2-m^2+i\epsilon}
	\frac{1}{[(p-k)^2 - m^2 +i\epsilon]^2} \\
	&+
	\frac{g^2}{\kappa^2}
	\int \frac{d^4k}{(2\pi)^4}
	f_{21}(k)
	\frac{1}{k^2-m^2+i\epsilon}
	\frac{1}{[(p-k)^2 - m^2 +i\epsilon]^3} \label{f21-cont}\\
	&+
	\frac{g^2}{\kappa^2}
	\int \frac{d^4k}{(2\pi)^4}
	f_{22}(k)
	\frac{1}{k^2-m^2+i\epsilon}
	\frac{1}{[(p-k)^2 - m^2 +i\epsilon]^2},\label{f22-cont}
\end{align}
where
\begin{align}\label{f1}
	f_1(k)
	=
	\frac{1}{4}
	\left[\Delta_1[(S(k)\oplus p)^2]
	+
	\Delta_1[(p \oplus S(k))^2]
	+
	\widehat{\Delta}_1[(k \oplus p)^2]
	+
	\widehat{\Delta}_1[(p \oplus k)^2]\right]
\end{align}
\begin{align}\label{f21}
	f_{21}(k)
	=
	\frac{1}{4}
	\left[(\Delta_1[(S(k)\oplus p)^2])^2
	+
	(\Delta_1[(p \oplus S(k))^2])^2
	+
	(\widehat{\Delta}_1[(k \oplus p)^2])^2
	+
	(\widehat{\Delta}_1[(p \oplus k)^2])^2\right]
\end{align}
\begin{align}\label{f22}
	f_{22}(k)
	=
	\frac{1}{4}
	\left[\Delta_2[(S(k)\oplus p)^2]
	+
	\Delta_2[(p \oplus S(k))^2]
	+
	\widehat{\Delta}_2[(k \oplus p)^2]
	+
	\widehat{\Delta}_2[(p \oplus k)^2]\right].
\end{align}

If we instead consider the case of one incoming particle with momentum $-p$ (to be more precise, the momentum $-p$ here needs to be interpreted as $p^*$ in \cite{Arzano:2020jro}), then all the reasoning is the same, except that in eq. \eqref{p1}, \eqref{p3} one must switch $p \mapsto -S(p)$ (or, to use the notation in \cite{Arzano:2020jro}, we perform the switch $p \mapsto S(p^*)$). One therefore obtains an equivalent expression to eq. \eqref{defamplitude}, but with numerators given by 
\begin{align}\label{g1}
	g_1(k)
	&=
	\frac{1}{4}
	\big[\Delta_1[(S(k)\oplus [-S(p)])^2]
	+
	\Delta_1[([-S(p)] \oplus S(k))^2] \nonumber \\
	&+
	\widehat{\Delta}_1[(k \oplus [-S(p)])^2]
	+
	\widehat{\Delta}_1[([-S(p)] \oplus k)^2]\big]
\end{align}
\begin{align}\label{g21}
	g_{21}(k)
	&=
	\frac{1}{4}
	\big[(\Delta_1[(S(k)\oplus [-S(p)])^2])^2
	+
	(\Delta_1[([-S(p)] \oplus S(k))^2])^2 \nonumber \\
	&+
	(\widehat{\Delta}_1[(k \oplus [-S(p)])^2])^2
	+
	(\widehat{\Delta}_1[([-S(p)] \oplus k)^2])^2\big]
\end{align}
\begin{align}\label{g22}
	g_{22}(k)
	&=
	\frac{1}{4}
	\big[\Delta_2[(S(k)\oplus [-S(p)])^2]
	+
	\Delta_2[([-S(p)] \oplus S(k))^2] \nonumber \\
	&+
	\widehat{\Delta}_2[(k \oplus [-S(p)])^2]
	+
	\widehat{\Delta}_2[([-S(p)] \oplus k)^2]\big].
\end{align}
Despite the fact that the expansion of each of the $\Delta$ in this case is different from the previous case with incoming momentum $S(p)$, after summing their contributions one can show that 
\begin{align}\label{relations}
    g_{1}(k) = -f_1(k),
    \qquad
    g_{21}(k) = f_{21}(k),
    \qquad
    g_{22}(k) = f_{22}(k).
\end{align}

 {Notice that both the expressions for the $f$ functions and for the $g$ functions hold in a generic frame, and so does eq. \eqref{relations}. For simplicity, we have included their full expression in the appendix \ref{AA}. Furthermore, notice that the only physically relevant quantity is the average of the amplitude computed assuming an incoming particle with momentum $S(p)$, i.e. the one defined in eq. \eqref{defamplitude}, and the one computed assuming an incoming particle with momentum $-p$, which we can call $i\Pi^{(-p)}_{TOT}(p^2)$. 
\begin{align}
    i\Pi_{\text{full}}(p^2)
    =
    \frac{i\Pi_{TOT}(p^2) + i\Pi^{(-p)}_{TOT}(p^2)}{2}
\end{align}
This is due to the fact that, upon noticing an incoming particle, we have no way to know whether its momentum is $S(p)$ or $-p$ (both of which describe an incoming particle). In light of eq. \eqref{relations}, it is therefore clear that the full amplitude (in a general frame) will not contain $O(1/\kappa)$ contributions since $g_1+f_1 = 0$, and the first non-trivial correction is at $O(1/\kappa^2)$. In section \ref{comframe}, we will specialize the general expressions in eq. \eqref{f1-a}, \eqref{f21-a}, \eqref{f22-a} to the c.o.m. frame in order to simplify them, and to perform explicit computations.}

 {Coming back to the singularity structure, }we see that in general, after expanding in powers of $1/\kappa$, we are left with integrands whose singularity structure is similar to the undeformed case, with the only difference being that some of the propagators have exponents different from one. We now present an argument which shows the validity of the Cutkosky cutting rules for each term in the expanded amplitude above, and we then reduce the computation of the cuts for the amplitude $i\Pi_{TOT}$ in eq. \eqref{defamplitude} to simpler amplitudes.

\section{Cutkosky rule in the deformed context}\label{cutdef-dim}
The Cutkosky rules by definition work on non-deformed QFT. Notice that in the non-deformed context they hold even in the presence of a cut-off if the cut propagators are still degenerate in the integration domain (and therefore if the argument of the respective Dirac delta after the cut can still go to zero in the integration domain). Indeed, the imaginary part does not depend on the cut-off (recall that counterterms coming from a Hermitian Lagrangian can only be real). 
Therefore, even in the presence of a cut-off, the imaginary part of a diagram can be computed by considering all the possible cuts.

We now show that the Cutkosky rules also correctly reproduce the singularity structure of each term in the $1/\kappa$ expansion of a $\kappa$-deformed amplitude in a  {$g\phi \star \varphi \star \varphi$ interaction model (as we showed above, all other possibilities of the star product give rise to the same amplitudes)}. The key point of the argument below lies in the fact that in the deformed case, one needs to introduce a cut-off to justify in a clearer way the expansion in powers of $1/\kappa$, so that each momentum is bounded by $\kappa$. Contrary to the non-deformed case, however, where one is free to move the value of the cut-off $\Lambda$ so that all the propagators can go on-shell, in the deformed context this is not true in general, since the cut-off cannot go above $\kappa$ to have a well-defined expansion in powers of $1/\kappa$.

Using this fact, we now consider a non-deformed model with interactions $g\tilde{\phi}\tilde{\varphi}^2$ and an additional interaction $G\Tilde{\varphi}\Tilde{\Phi}^2$ where $\tilde{\Phi}$ has a mass $M$ such that $M/\kappa >1$.  {The $\,\, \tilde{} \,\,$ on top of the fields indicate that these are non-deformed fields. Notice that, since we are in the non-deformed case, there is now no order ambiguity in both vertices. The aim of this section is to show that, if one focuses on just the singularity structure of 1-loop amplitudes, then the terms in the $1/\kappa$ expansion of a deformed model containing two deformed fields ($\phi$, $\varphi$) with $g\phi \star \varphi \star \varphi$ interaction vertex, are equivalent to specific higher-order loop corrections in a non-deformed model with three non-deformed scalar fields ($\tilde{\phi}, \tilde{\varphi}$, and the additional auxiliary heavy field $\tilde{\Phi}$), with interaction vertices $g\tilde{\phi}\tilde{\varphi}^2$ and $G\tilde{\varphi}\tilde{\Phi}^2$. As such, since the Cutkosky rules hold in the non-deformed context, we extend their validity to the deformed amplitude, expanded in powers of the deformation parameter $1/\kappa$. In what follows, we allow for arbitrary numerators which do not influence the singularity structure.} We concentrate on 1-loop amplitudes, but the reasoning can be generalized to more than one loop. Consider the following diagram.
\begin{figure}[h]
    \centering
    \includegraphics[width=0.25\linewidth]{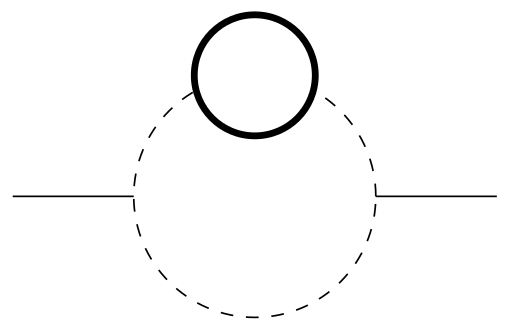}
    \label{fig:2}
\end{figure}

The bold lines represent propagators containing the mass $M$. Its amplitude (with external legs excluded) reads
\begin{align}
	i\Pi
	=
	(ig)^2(iG)^2
	\int dk \, dl \,\,
	N(k,l,p,m,M)
	&\frac{i}{k^2-m^2 + i\epsilon} 
	\frac{i}{[(p-k)^2-m^2 + i\epsilon]^2} \times \nonumber \\
	&\times
    \frac{i}{l^2-M^2 + i\epsilon}
	\frac{i}{(l-k)^2-M^2 + i\epsilon}
\end{align}
where all loop momenta are limited in modulus by $\kappa$, and $N$ is some analytic function in the masses, loop momenta, and external momenta.

There are only two types of cut, represented by the vertical lines in the figure below.
\begin{figure}[h]
    \centering
    \includegraphics[width=0.25\linewidth]{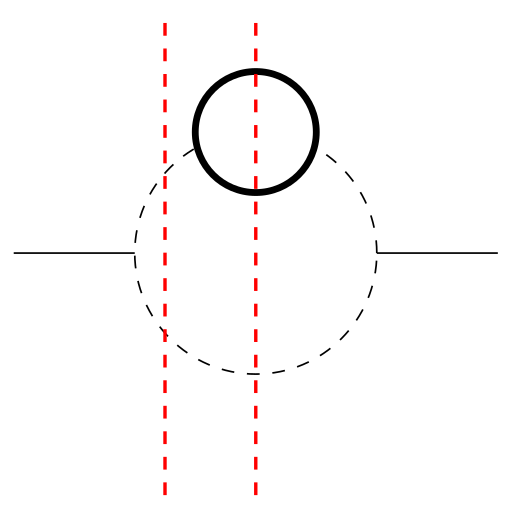}
    \label{fig:3}
\end{figure}
Since we are considering the case where the mass $M$ of the bold loop is above the cut-off scale, the cutting rules still continue to hold, but this time the cut passing through the bold loop contributes as zero. Indeed, none of its propagators can go on-shell, so that the Dirac delta associated to the cut is identically zero. This is reflected in the full amplitude, where the two propagators no longer generate singularities and can be absorbed in the analytic numerator. The imaginary part is therefore only due to the propagators of the outer loop. Therefore, the Cutkosky cutting rules still capture correctly the singularity structure of deformed diagrams. 

The above diagram only provides justification for the first term in the $1/\kappa$ expansion of a 1-loop deformed diagram in terms of a 2-loop non-deformed diagram. In general, however, higher order coefficients in the expansion of a $\kappa$-deformed amplitude in powers of $1/\kappa$ can contain one propagator being exponentiated to some integer power $n$, and the second one to the power $m$. In this case, one can reproduce the same reasoning as above, relating this deformed diagram to the non-deformed one represented in the picture below.
\begin{figure}[h]
    \centering
    \includegraphics[width=0.3\linewidth]{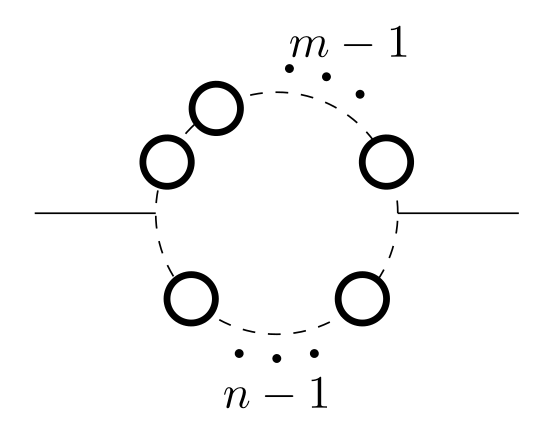}
    \label{fig:4}
\end{figure}
The top side contains $m-1$ additional heavy loops, and the lower one $n-1$ additional heavy loops. Reasoning as before, and setting $M/\kappa > 1$, shows that Cutkosky rules also reproduce the singularity structure of this diagram. Since this holds for any $m$ and $n$, it holds at any order in perturbation in powers of $1/\kappa$. 

Although the above argument is useful to establish the validity of the cutting rules, its usefulness is limited in actual computations. To avoid this, we now introduce a procedure that shows in a different way the fact that the cutting rules reproduce the singularity structure of the expanded $\kappa$-deformed amplitude. The key point is to reduce each term of the expansion of the deformed amplitude to simpler terms, using algebraic and analytic identities, such that the validity of the Cutkosky rules is again manifest. Furthermore, using these simpler terms, the actual computations in concrete cases is much more simplified. We limit our procedure to the second order in the expansion in $1/\kappa$, but the reasoning can be further generalized to higher orders. 

\section{Simplified Cutkosky rule in the deformed context}\label{cutdef}

Consider the case of a general, non-trivial numerator $f(k,p,m)$, so that we have 
\begin{align}
	i\widetilde{\Pi}_d(p^2)
	&=
	(ig)^2
	\int \frac{d^4 k}{(2\pi)^4} \,\, f(k,p,m)
	\frac{i}{k^2 - m^2 +i\epsilon}
	\frac{i}{[(p-k)^2 - m^2 +i\epsilon]^2} \\
	&=
	(ig)^2
	\int \frac{d^4 k}{(2\pi)^4} \,\, f(k,p,m)
	\frac{i}{k^2 - m^2 +i\epsilon}
	\left(
	\frac{\partial}{\partial m^2}
	\frac{i}{(p-k)^2 - m^2 +i\epsilon}
	\right).
\end{align}
Recall that any function can be written as the sum of its even and odd part under exchange $k\mapsto p-k$ as follows (we are still considering functions which also depend on $p$ and $m$, but for simplicity we keep these additional dependence implicit)
\begin{align}\label{EOdef}
	f(k)
	=
	\underbrace{\frac{f(k) + f(p-k)}{2}}_{E(k)}
	+
	\underbrace{\frac{f(k) - f(p-k)}{2}}_{O(k)}
\end{align}
\begin{align}
	E(k)
	\xrightarrow{k\rightarrow p-k}
	E(p-k) = E(k),
	\qquad
	O(k)
	\xrightarrow{k\rightarrow p-k}
	O(p-k) = -O(k).
\end{align}
Let us also use the convention
\begin{align}
	A = (p-k)^2 - m^2 +i\epsilon,
	\qquad
	B=k^2 - m^2 +i\epsilon.
\end{align}
It is clear that 
\begin{align}
	\frac{1}{A^2} = \frac{\partial}{\partial m^2} \frac{1}{A},
	\qquad
	\frac{1}{B^2} = \frac{\partial}{\partial m^2} \frac{1}{B}.
\end{align}
Integrating by parts one can show that
\begin{align}\label{evenpart}
	g^2
	\int \frac{d^4k}{(2\pi)^4}
	\frac{E(k)}{A^2B}
	=
	\frac{g^2}{2}
	\frac{\partial}{\partial m^2}
	\int \frac{d^4k}{(2\pi)^4}
	\frac{f(k)}{AB}
	-
	\frac{g^2}{2}
	\int \frac{d^4k}{(2\pi)^4}
	\frac{\partial f(k)}{\partial m^2}
	\frac{1}{AB}.
\end{align}
We are only left with the need to evaluate the odd part of the integral, i.e.
\begin{align}
	g^2
	\int \frac{d^4k}{(2\pi)^4}
	\frac{O(k)}{A^2B}.
\end{align}
To simplify the reasoning, we first change variables $k \mapsto \frac{p+k}{2}$. Under this change of coordinates, the symmetry properties under $k \mapsto p-k$ send $\frac{p+k}{2} \mapsto \frac{p-k}{2}$, and therefore $k\mapsto -k$. Terms odd (even) under exchange $\kappa \mapsto p-k$ are now odd (even) under $k\mapsto -k$. The new integrand is therefore 
\begin{align}
	\frac{g^2}{2^4}
	\int \frac{d^4k}{(2\pi)^4}
	\frac{\widehat{O}(k)}{\widehat{A}^2\widehat{B}}.
\end{align}
We now notice that $\widehat{O}(k)$ is a scalar (there are no free indices). It is also assumed to be polynomial in the loop and external momenta. Therefore one must have $
	\widehat{O}(k)
	=
	(pk)
	\widehat{E}(k) 
$
where $\widehat{E}(k)$ is a n even function under $k \mapsto -k$. We now use the identity 
\begin{align}
	\frac{pk}{\widehat{A}^2 \widehat{B}}
	=
	\frac{1}{\widehat{A}^2}
	-
	\frac{1}{\widehat{A}\widehat{B}}. 
\end{align}
One can rewrite the integral containing only $A^2$ in the denominator in terms of derivative of integrals with only $A$, which cannot have any imaginary part since they are tree level. Therefore, since our aim is to apply the Cutkosky rule to compute the imaginary part, we can ignore those contributions.  To restore the symmetry of the form $k \mapsto p-k$, it is sufficient to change variables once again with $k \mapsto 2k - p$. In other words (keeping in mind that we are only interested in the imaginary part) we have
\begin{align}\label{A2B}
	g^2
		\int \frac{d^4k}{(2\pi)^4}
		\frac{f(k)}{A^2B}
		=
		\frac{g^2}{2}
		\frac{\partial}{\partial m^2}
		\int \frac{d^4k}{(2\pi)^4}
		\frac{f(k)}{AB}
		-
		\frac{g^2}{2}
		\int \frac{d^4k}{(2\pi)^4}
		\left(
		\frac{\partial f(k)}{\partial m^2}
		+ 
		2
		\frac{O(k)}{2pk - p^2}
		\right)
		\frac{1}{AB}.
\end{align}
Notice that, in light of the computations in section \ref{nondefloopcorr}, all the quantities in the RHS are well defined.

One can proceed along the same line to obtain a similar formula for the case in which the denominator is of the form $A^3B$. In fact, using the same short-hand notation, we have
\begin{align}
	i\widetilde{\Pi}_d(p^2)
	=
	g^2
	\int 
	\frac{d^4k}{(2\pi)^4}
	\frac{f(k)}{A^3B}
\end{align}
and we will indicate $\partial := \frac{\partial}{\partial m^2}$ for brevity. We have
\begin{align}\label{provv1}
	g^2
	\int 
	\frac{d^4k}{(2\pi)^4}
	\frac{f(k)}{A^3B}
	=
	\frac{g^2}{2}
	\partial 
	\int 
	\frac{d^4k}{(2\pi)^4}
	\frac{f(k)}{A^2B}
	-
	\frac{g^2}{2}
	\int 
	\frac{d^4k}{(2\pi)^4}
	\frac{\partial f(k)}{A^2B}
	-
	\frac{g^2}{2} 
	\int 
	\frac{d^4k}{(2\pi)^4}
	\frac{f(k)}{A^2B^2}.
\end{align}
Thanks to eq. \eqref{A2B} we know how to compute the first two terms. For the last term, one can notice that 
\begin{align}
	-2\int 
	\frac{d^4k}{(2\pi)^4}
	\frac{f(k)}{A^2B^2}
	&=
	-
	\partial^2
	\int 
	\frac{d^4k}{(2\pi)^4}
	\frac{f(k)}{AB}
	+
	\int 
	\frac{d^4k}{(2\pi)^4}
	\frac{\partial^2 f(k)}{AB}
	+
	2
	\int 
	\frac{d^4k}{(2\pi)^4}
	\frac{\partial E(k)}{A^2B} 
    +
	\int 
	\frac{d^4k}{(2\pi)^4}
	\frac{E(k)}{A^3B}.
\end{align}
Multiplying both sides by $1/4$ and recalling the definition of even part in eq. \eqref{EOdef} one can substitute the resulting expression into eq. \eqref{provv1}. The resulting terms can now all be evaluated, with the exception of a remaining term with integrand $E/A^3B$. This can however be transferred to the LHS and switched for an odd term using the fact that if $f = C + E/2$, then $f= C-O/2$ (where $f=E+O$ and $C$ represents other terms). Again notice that we can repeat the reasoning of the previous case, since we can write
\begin{align}
	\frac{O(k)}{A^3B}
	=
	\frac{O(k)}{2pk - p^2}
	\left(
	\frac{1}{A^3}
	-
	\frac{1}{A^2B}
	\right).
\end{align}
Once again, we can ignore the first term because it does not have any imaginary part, since it can be written in terms of derivatives of tree diagrams. Therefore, we have the final formula (which we recall is valid for the computation of the imaginary part)
\begin{align}\label{A3B}
	g^2
	\int 
	\frac{d^4k}{(2\pi)^4}
	\frac{f(k)}{A^3B}
	&=
	g^2
	\partial 
	\int 
	\frac{d^4k}{(2\pi)^4}
	\frac{f(k)}{A^2B}
	-
	\frac{g^2}{2}
	\partial^2
	\int 
	\frac{d^4k}{(2\pi)^4}
	\frac{f(k)}{AB} 
	+
	\frac{g^2}{2}
	\int 
	\frac{d^4k}{(2\pi)^4}
	\frac{\partial^2 f(k)}{AB}\nonumber \\
    &+
	g^2
	\int 
	\frac{d^4k}{(2\pi)^4}
	\left(
	-
	\partial f(k)
	+
	2\partial E(k)
	+
	\frac{O(k)}{2pk - p^2}
	\right)
	\frac{1}{A^2B}.
\end{align}
Notice that in each of the terms in the RHS, the singularity structure is simpler than the one in the LHS, and in particular is exactly the same singularity structure of a 1-loop non-deformed correction to the propagator (with a different numerator, which however does not influence the singulariries). The validity of the Cutkosky rules is therefore once again manifest. We now first show an example of an explicit check of the validity of eq. \eqref{A2B} (which highlights how much simpler are the computations using the cutting rules), and we then use both identities above to compute the full imaginary part of the $\kappa$-deformed 1-loop correction up to second order in $1/\kappa$. 

\subsection{Example of checking the deformed Cutkosky rules}\label{checkdefcut}

We can show here a quick check of eq. \eqref{A2B} in the simple case where the numerator is given by $k^2$. We start from 
\begin{align}
	i\Pi_d(p^2)
	=
	(ig)^2
	\int \frac{d^4 k}{(2\pi)^4} 
	k^2
	\underbrace{\frac{i}{k^2 - m^2 +i\epsilon}}_{B}
	\underbrace{\frac{i}{[(p-k)^2 - m^2 +i\epsilon]^2}}_{A}.
\end{align}
This time the numerator has an even and an odd part under the exchange $k \mapsto p-k$
\begin{align}
	E(k)
	=
	\frac{k^2 + (p-k)^2}{2},
	\qquad
	O(k)
	=
	\frac{k^2 - (p-k)^2}{2}
    =
	\frac{1}{2}
	(2pk - p^2),
\end{align}
so that the above formula in eq. \eqref{A2B} simplifies to 
\begin{align}
	g^2
	\int \frac{d^4k}{(2\pi)^4}
	\frac{k^2}{A^2B}
	=
	\frac{g^2}{2}
	\frac{\partial}{\partial m^2}
	\int \frac{d^4k}{(2\pi)^4}
	\frac{k^2}{AB}
	-
	\frac{g^2}{2}
	\int \frac{d^4k}{(2\pi)^4}
	\frac{1}{AB}.
\end{align}
The second term is exactly as the non-deformed case, with an additional $-\frac{1}{2}$, so that using eq. \eqref{rescut1} we get
\begin{align}\label{st}
	\Im
	\left(
	-
	\frac{g^2}{2}
	\int \frac{d^4k}{(2\pi)^4}
	\frac{1}{AB}
	\right)
	=
	-\frac{1}{2}
	\frac{g^2}{16\pi}
	\sqrt{1-4\frac{m^2}{p^2}}.
\end{align}
For the first term, if we ignore the derivative we see that this is just the same as eq. \eqref{prelim} with an additional $m^2/2$ in front, so that 
\begin{align}\label{ft}
	\Im [iI]
	=
	\frac{m^2}{2}
	\frac{g^2}{16\pi}
	\sqrt{1-4\frac{m^2}{p^2}}.
\end{align}
Taking the first derivative of eq. \eqref{ft} with respect to $m^2$ and adding it to eq. \eqref{st} we get
\begin{align}\label{cutcheck}
	\Im(i\Pi_d(p^2))
	=
	-
	\frac{g^2}{16\pi}
	\frac{m^2}{p^2}
	\left(
	\sqrt{1-4\frac{m^2}{p^2}} 
	\right)^{-1}
\end{align}
because the other two contributions cancel each other out.

To compute the imaginary part directly, we need to merge the denominators using the Feynman trick, and then perform both the shift $k \mapsto k + p(1-x)$ and perform the Wick rotation. After these canonical steps we are left with
\begin{align}
	i\Pi(p^2)
	=
	-ig^2
	\int_0^1 dx
	\,\, 2(1-x)
	\int
	\frac{d^4 k}{(2\pi)^4}
	\frac{-k_E^2 + p^2 (1-x)^2}{(k_E^2 + \Delta)^3}.
\end{align}
We can compute the two integrals separately, starting with the one with $k_E^2$ at the numerator. We call this integral $I_0$. The momentum space integral can be computed with eq. \eqref{dimregint} with $a=1$ and $b=3$. Therefore the momentum space integral is again given by eq. \eqref{dimregint}, so we can follow the same reasoning, and we end up with 
\begin{align}
	i\Pi(p^2)
	=
	-i\frac{g^2}{(4\pi)^2}
	\int_0^1 dx
	\,\, 2(1-x)
	\log(m^2 - p^2 x(1-x)).
\end{align}
By changing $x \mapsto 1-x$ one can show that 
\begin{align}
	-i\frac{g^2}{(4\pi)^2}
	\int_0^1 dx
	\,\, 2x
	\log(m^2 - p^2 x(1-x))
	=
	-i\frac{g^2}{(4\pi)^2}
	\int_0^1 dx
	\log(m^2 - p^2 x(1-x))
\end{align}
and this is the same integral as the one encountered in section \ref{nondefloopcorr}. Therefore the imaginary part of the term with $k_E^2$ in the numerator is
\begin{align}
	\Im(I_0)
	=
	+
	\frac{g^2}{16\pi}
	\sqrt{1-4\frac{m^2}{p^2}}.
\end{align}
The term with $p^2$ at the numerator instead is given by $a=0$ and $b=3$ in eq. \eqref{dimregint} so that we need to compute the imaginary part of
\begin{align}
	-i\frac{g^2}{32\pi^2}
	\int_0^1 dx
	\,\, 2x^3
	\frac{p^2 }{m^2 - p^2 x(1-x)}
\end{align}
where we switched $x\mapsto 1-x$ to simplify the expression. Using the notation
\begin{align}
	I_1 := 
	-i\frac{g^2}{32\pi^2}
	\int_0^1 dx
	\,\, 2x^3
	\frac{p^2 }{m^2 - p^2 x(1-x)},
\qquad
	I_2 := 
	-i\frac{g^2}{32\pi^2}
	\int_0^1 dx
	\,\,
	\frac{2x}{m^2 - p^2 x(1-x)},
\end{align}
\begin{align}
	I_3 :=
	-i\frac{g^2}{32\pi^2}
	\int_0^1 dx
	2\log(m^2 - p^2 x(1-x))
	+
	4
	-
	2\log(m^2),
\end{align}
one can show that 
\begin{align}
	\Im(I_1)
	=
	-\Im(I_3)
	+
	m^2 \Im(I_2)
\end{align}
where
\begin{align}
	\Im(I_3) 
	=
	+\frac{g^2}{16\pi}
	\sqrt{1-4\frac{m^2}{p^2}}
\qquad
	\Im(I_2)
	=
	-\frac{g^2}{16\pi}
	\frac{1}{p^2}
	\left(
	\sqrt{1-4\frac{m^2}{p^2}}
	\right)^{-1}.
\end{align}
Adding all the contributions results in 
\begin{align}
	\Im(I_0)
	+
	\Im(I_1)
	=
	-\frac{g^2}{16\pi}
	\frac{m^2}{p^2}
	\left(
	\sqrt{1-4\frac{m^2}{p^2}}
	\right)^{-1}
\end{align}
which is exactly the same result obtained in eq. \eqref{cutcheck} with the Cutkosky rule, as expected.

\section{$\kappa$-deformed imaginary part of the 1-loop correction}\label{imcomp}

We now apply our relations \eqref{A2B} and \eqref{A3B} to compute the full imaginary part of the 1-loop correction to the propagator, up to second order in $1/\kappa$. In the non-deformed case, one could perform the computations in the centre of mass (c.o.m.) frame because of the trivial numerator. In the deformed case, we will see that the numerator (after the expansion in powers of $1/\kappa$) is not a Lorentz scalar in general, and therefore one should perform the computations in an arbitrary frame in order to get the most general result. 

Here, however, we will perform the full computations in the c.o.m.. Doing so, we will see that the deformed contributions is proportional to $p^2/\kappa^2$. We will then argue that the full result in an arbitrary frame must be of the same type, containing an additional term proportional $p_0^2/\kappa^2$.

In order to get the full amplitude, we will take the average of the two amplitudes where the incoming momentum is $S(p)$ and $-p$. In fact, from a phenomenological point of view, we can only see a particle moving towards us at a certain speed, and we do not know whether its momentum is $-p$ or $S(p)$. 

\subsection{Centre of mass frame}\label{comframe}

In the c.o.m. frame  {equations \eqref{f1-a}, \eqref{f21-a}, \eqref{f22-a} reduce to}
\begin{align}
	f_1(k)
	=
	0,
\end{align}
\begin{align}
	f_{21}(k)
	=
	2 \left(k_1^2 p_0+k_2^2 p_0+k_3^2 p_0\right)^2,
\end{align}
\begin{align}
	f_{22}(k)
	=
	\frac{1}{4} \left(2 k_0^3 p_0+k_0^2 \left(k_1^2+k_2^2+k_3^2-4 p_0^2\right)+2 k_0 p_0^3+k_1^2 p_0^2+k_2^2 p_0^2+k_3^2 p_0^2\right),
\end{align}
 {so that, as expected,} we only have the $1/\kappa^2$ contribution to the amplitude. Since there is no explicit dependence on the mass $m$, we can use the simplified relations
\begin{align}\label{A2B-loop-c.o.m.}
	g^2
		\int \frac{d^4k}{(2\pi)^4}
		\frac{f(k)}{A^2B}
		=
		\frac{g^2}{2}
		\frac{\partial}{\partial m^2}
		\int \frac{d^4k}{(2\pi)^4}
		\frac{f(k)}{AB}
		-
		g^2
		\int \frac{d^4k}{(2\pi)^4}
		\left(
		\frac{O(k)}{2pk - p^2}
		\right)
		\frac{1}{AB},
\end{align}
\begin{align}\label{A3B-loop-c.o.m.}
	g^2
		\int 
		\frac{d^4k}{(2\pi)^4}
		\frac{f(k)}{A^3B}
		=
		g^2
		\partial 
		\int 
		\frac{d^4k}{(2\pi)^4}
		\frac{f(k)}{A^2B}
		-
		\frac{g^2}{2}
		\partial^2
		\int 
		\frac{d^4k}{(2\pi)^4}
		\frac{f(k)}{AB}
		+
		g^2
		\int 
		\frac{d^4k}{(2\pi)^4}
		\left(
		\frac{O(k)}{2pk - p^2}
		\right)
		\frac{1}{A^2B}.
\end{align}
We start with the computations of eq. \eqref{A2B-loop-c.o.m.}, since we will need it for the computation of eq. \eqref{A3B-loop-c.o.m.}. The starting point is the formula 
\begin{align}
	2\Im [iA]
	=
	\frac{g^2}{16\pi^2}
	\int d^3 k 
	\frac{N(k_0, k_0+q_0, \mathbf{k}, 0)}{k_0 q_0}
	\delta(q_0 + k_0  - p_0).
\end{align}
where $N$ stands for a generic numerator with $\mathbf{p}=0$, and proceeding as before in section \ref{nondefloopcorr} we have $p_\mu = (p_0, 0,0,0)$, $k_\mu = (\sqrt{|\mathbf{k}|^2 + m^2}, \mathbf{k})$, and $q_\mu = (\sqrt{|\mathbf{k}|^2 + m^2}, -\mathbf{k})$, so that the above integral becomes (we also pass to polar coordinates in momentum space)
\begin{align}
	2\Im [iA]
	&=
	\frac{g^2}{32\pi^2}
	\int d|\mathbf{k}| \,\sin\theta d\theta\, d\phi\, 
	N(\sqrt{|\mathbf{k}|^2+ m^2}, 2\sqrt{|\mathbf{k}|^2+ m^2}, |\mathbf{k}|,\theta,\phi, 0) \times \nonumber \\
    &\times
	\frac{|\mathbf{k}|}{\sqrt{|\mathbf{k}|^2+ m^2}}
	\delta\left(|\mathbf{k}| - \sqrt{\frac{p_0^2}{4} - m^2}\right).
\end{align}
Using this relation, we can compute the imaginary part of the terms in eq. \eqref{f21-cont} and \eqref{f22-cont} in the amplitude. 

Starting from eq. \eqref{f21-cont}, one can show that the first two terms in the RHS of eq. \eqref{A3B-loop-c.o.m.} are equal and opposite, while the third is zero because $O_{21} = 0$ identically in the c.o.m. frame. Therefore, we get that the contribution of eq. \eqref{f21-cont} to the amplitude vanishes
\begin{align}
	\int 
	\frac{d^4k}{(2\pi)^4}
	\frac{f_{21}(k)}{A^3B}
	=
	0.
\end{align}

We can therefore compute the contribution given by eq. \eqref{f22-cont}. One can show that the contributions to the RHS of eq. \eqref{A2B-loop-c.o.m.} are given by 
\begin{align}\label{cut1-bis}
	\Im
    \left(
    \frac{g^2}{2}
		\frac{\partial}{\partial m^2}
		\int \frac{d^4k}{(2\pi)^4}
		\frac{f_{22}(k)}{AB}
    \right)
		=
		\frac{g^2 \left(60 m^2 p_0-19 p_0^3\right)}{1024 \pi \sqrt{p_0^2-4 m^2}},
\end{align}
\begin{align}\label{cut2-bis}
	\Im
    \left(
    -
	g^2
	\int \frac{d^4k}{(2\pi)^4}
	\left(
	\frac{O(k)}{2pk - p^2}
	\right)
	\frac{1}{AB}
    \right)
	=
	-\frac{g^2 \left(p_0^2-4 m^2\right)^{3/2}}{512 \pi p_0}.
\end{align}
Summing the results of eq. \eqref{cut1-bis} and  \eqref{cut2-bis} we have the full result of the deformed contribution to the imaginary part of the 1-loop correction to the scalar propagator. One can express the full deformed imaginary part $\Im[i\Pi_{TOT}(p^2)]$ in terms of the non deformed one $\Im[i\Pi_{U}(p^2)]$ (whose expression can be found in eq. \eqref{rescut1}) as follows
\begin{align}
	\Im [i\Pi_{TOT}(p^2)]
	=
	\Im [i\Pi_U(p^2)]
	\left(
	1
	+
	\frac{1}{\kappa^2}
	\frac{32 m^4-76 m^2 p^2+21 p^4}{256 m^2-64 p^2}
	\right).
\end{align}
Notice that we have substituted $p_0$ with $p$ since in the c.o.m. they are the same.
This expression can be further simplified with some algebra, arriving at the final result 
\begin{align}
	\Im [i\Pi_{TOT}(p^2)]
	=
	\Im [i\Pi_U(p^2)]
	\left[
	1
	-
	\frac{p^2}{\kappa^2}
	\left(
	\frac{512\pi^2}{g^4}
	(\Im [i\Pi_U(p^2)])^2
	+
	\frac{g^4}{64\pi^2(\Im [i\Pi_U(p^2)])^2}
	+
	\frac{1}{8}
	\right)
	\right].
\end{align}
In light of eq. \eqref{relations}, the same result holds for an incoming particle of momentum $-p$, and therefore the above result is also true for the average of the two cases.

\subsection{General frame and decay widths}\label{generalframe}

 {As already noticed in section \ref{defamp},} we do not get lower order corrections moving away from the c.o.m. frame, since because of eq. \eqref{relations} we have that the $1/\kappa$ contribution identically vanishes in any frame when taking the average. The only difference with the c.o.m. frame is the presence of additional terms in the numerator proportional to $p_0$ and $\mathbf{p} = \sqrt{p^2 - p_0^2}$. Therefore, for dimensional reasons, the general correction can only have the form 
\begin{align}
    \Im [i\Pi_{TOT}(p^2)]
	=
	\Im [i\Pi_U(p^2)]
	\left[
	1
    +
    \alpha \frac{p^2}{\kappa^2}
    +
    \beta \frac{p_0^2}{\kappa^2}
    \right],
\end{align}
where $\alpha$ and $\beta$ are coefficients whose computation is beside the scope of the present work, although (ignoring the boost factors $\gamma$) we expect them to be of the same order of magnitude as the coefficients in the c.o.m. frame. To understand in detail their dependence on the boost factor $\gamma$ would require their full computation. For single diagrams, the $\alpha$ and $\beta$ coefficients will be in general different, but as we saw above, if we sum all the contributions then $\alpha$ and $\beta$ are the same, regardless of whether the incoming particle has momentum $S(p)$ or $-p$. 

This allows us to put some first stringent limits on the behaviour of the decay width of unstable particles in the model described in \cite{Bevilacqua:2023pqz, Arzano:2020jro, Bevilacqua:2022fbz}. Indeed, by using the Dyson resummation for the above 1-loop correction, we get
\begin{align}
    i\Delta
    =
    \frac{1}{p^2 - m^2}
    \sum_{i=0}^\infty
    \left(-
    \frac{\Pi_{TOT}(p^2)}{p^2-m^2}
    \right)
    =
    \frac{1}{p^2-m^2 + \Pi_{TOT}(p^2)},
\end{align}
where the quantity $\Pi_{TOT}(p^2)$ will contain both a real and an imaginary part. The real part is finite, since the zeroth order in $1/\kappa$ is the finite non-deformed term, and each subsequent integral in perturbation series has a cut-off at $\kappa$. The imaginary part is the one we computed above. The relation 
\begin{align}
    m\Gamma
    =
    \Im\left(\Pi_{TOT}(m^2)\right)
\end{align}
relates the decay width to the imaginary part of the 1-loop correction. One immediately sees, therefore, that the 1-loop contributions to the decay width are at best of the order of $p_0^2/\kappa^2$, and therefore negligible.

\section{Conclusions and discussion} \label{conclusion}

In this paper we showed the validity of the Cutkosky cutting rules order by order in the expansion of a deformed amplitude (for definiteness, and since our discussion is fairly general, we call the deformation parameter $\lambda$). In general, after expansion in $\lambda$, these amplitudes will have a denominator with propagators where canonical conservation of momentum is applied, but with either (or both) propagators exponentiated to a power different than one. We showed that by truncating the integration domain (a requirement needed to have a well-defined expansion in $\lambda$), the Cutkosky rules still reproduce correctly the singularity structure of the terms in the expansion of the deformed amplitude. The argument is based on relating each term in the expansion in $\lambda$ to non-deformed amplitude containing a particle of high mass. One could in principle apply directly the cutting rules and proceed with the computations. It is instead much simpler to first reduce the powers of the propagator in the denominator by using several algebraic and analytic identities. Although we showed how to do this directly only up to second order in $\lambda$, the reasoning can be easily generalized, and it provides a different way to show that indeed the cutting rules still capture the singularity structure of the expansion in $\lambda$ of the deformed amplitude. Pragmatically, using this approach one reduces the computation of the imaginary part to simpler amplitudes. Iterating the procedure in the same way until there are only terms with just canonical propagators, the application of the Cutkosky rules become trivial. 

Notice that the use of the cutting rules simplifies dramatically the computation of the imaginary part of loop corrections to the propagator. If one were to proceed with the canonical computations, one would first need to merge the denominator using the Feynman trick described by the relation
\begin{align}
	\frac{1}{A^a B^b}
	=
	\frac{\Gamma(a+b)}{\Gamma(a)\Gamma(b)}
	\int_0^1
	dx \,\,
	\frac{x^{a-1}(1-x)^{b-1}}{[Ax + B(1-x)]^{a+b}}.
\end{align}
After the shift in the loop momentum, one would first need to compute the momentum-space integrals, and then integrate in $x$. After this, one needs to extract the imaginary part from the result. Due to the fact that numerators become quickly very long after expansion in $\lambda$, this makes it very difficult to extract the imaginary part of even reasonably simple corrections, as was already noted in \cite{Bevilacqua:2023tre}. The approach presented in this paper makes for an easier computation of higher-order terms in the $\lambda$ expansion.

Due to this simplicity, we were able to show an example of calculation of the imaginary part of a diagram in a particular model. We chose this model also because, in previous publications \cite{Bevilacqua:2022fbz}, it was qualitatively argued that the first-order correction due to deformation of the imaginary part of the 1-loop diagram should be negligible. This was already partially checked in \cite{Bevilacqua:2023tre} where, for simplicity, only the imaginary part coming from the logarithms was studied. Here we confirmed in full generality that this is still the case when considering the full amplitude. This is an important result from a phenomenological point of view. Indeed, the fact that the decay width $\Gamma=1/\tau$ (where $\tau$ is the decay time of the unstable particle) is not affected by deformation up to negligible terms proportional to $1/\kappa^2$, makes it evident that the deformation effects described in \cite{Bevilacqua:2022fbz, Bevilacqua:2024jpy} are purely due to the behaviour under $CPT$ transformation of the fields.

From both an experimental and phenomenological point of view, the next natural step would be to extend the above computations to more loops. In this context, however, the non-planar graphs significantly complicate the matter. Furthermore, it would be interesting to consider in detail the implication of the above arguments about unitarity for general deformed amplitudes. We will address these interesting points in future publications.

\section*{Acknowledgements}

I would like to thank Jerzy Kowalski-Glikman and Wojciech Wi\'slicki for their guidance and their help, as well as for reading through the preliminary versions of this manuscript, providing helpful advice. I would also like to thank Alice Boldrin, for both her patience and her help in the writing of this paper. This work was supported by funds provided by the National Science Center, project no. 2019/33/B/ST2/00050.


\appendix
\section{ {Full expressions for $f_1$, $f_{21}$, $f_{22}$}}\label{AA}
 {
We here report the general result for the functions $f_1$, $f_{21}$, $f_{22}$ in eq. \eqref{f1}, \eqref{f21}, \eqref{f22}. Due to eq. \eqref{relations}, these are sufficient to determine $g_1$, $g_{21}$, $g_{22}$.
\begin{align}\label{f1-a}
	f_1(k)
	=
	-k_0 \left(p_1^2+p_2^2+p_3^2\right)+k_1 p_0 p_1+p_0 (k_2 p_2+k_3 p_3),
\end{align}
\begin{align}\label{f21-a}
	f_{21}(k)
	&=
	\left(-k_0 k_1 p_1-k_0 k_2 p_2-k_0 k_3 p_3+k_1^2 p_0+k_2^2 p_0+k_3^2 p_0\right)^2 \nonumber \\
	&+
	\left(-k_1 p_1 (k_0+p_0)-k_2 p_2 (k_0+p_0)+(k_3-p_3) (k_3 p_0-k_0 p_3)+k_0 \left(p_1^2+p_2^2\right)+k_1^2 p_0+k_2^2 p_0\right)^2 \nonumber \\
	&+
	\left(-k_0 \left(p_1^2+p_2^2+p_3^2\right)+k_1 p_0 p_1+p_0 (k_2 p_2+k_3 p_3)\right)^2,
\end{align}
\begin{align}\label{f22-a}
	f_{22}(k)
	&=
	\frac{1}{4} \Big[2 k_0^3 p_0+k_0^2 \left(k_1^2-3 k_1 p_1+k_2^2-3 k_2 p_2+k_3^2-3 k_3 p_3-4 p_0^2+p_1^2+p_2^2+p_3^2\right) \nonumber \\
	&+
	k_0 p_0 \left(3 (k_1 p_1+k_2 p_2+k_3 p_3)+2 p_0^2\right)+k_1^3 p_1+k_1^2 \left(p_2 (k_2-p_2)+p_3 (k_3-p_3)+p_0^2-3 p_1^2\right) \nonumber \\
	&+
	k_1 p_1 \left(k_2^2-4 k_2 p_2+k_3^2-4 k_3 p_3-3 p_0^2+p_1^2+p_2^2+p_3^2\right)+k_2^3 p_2 \nonumber \\
    &+
    k_3 p_3 \left(k_2^2-4 k_2 p_2+k_3^2-3 p_0^2+p_1^2+p_2^2\right) \nonumber \\
	&+
	k_2^2 p_0^2-k_2^2 p_1^2-3 k_2^2 p_2^2+p_3^2 \left(k_2 (p_2-k_2)-3 k_3^2+p_0^2\right) \nonumber \\
    &+
    k_2 k_3^2 p_2-3 k_2 p_0^2 p_2+k_2 p_1^2 p_2+k_2 p_2^3 \nonumber \\
	&+
	k_3^2 p_0^2-k_3^2 p_1^2-k_3^2 p_2^2+k_3 p_3^3+p_0^2 p_1^2+p_0^2 p_2^2\Big].
\end{align}
For simplicity, we also write down the even and odd parts. We have
\begin{align}
	E_1(k)
	=
	0,
\end{align}
\begin{align}
	O_1(k)
	=
	-k_0 \left(p_1^2+p_2^2+p_3^2\right)+k_1 p_0 p_1+p_0 (k_2 p_2+k_3 p_3),
\end{align}
\begin{align}
	E_{21}(k)
	&=
	\left(-k_0 k_1 p_1-k_0 k_2 p_2-k_0 k_3 p_3+k_1^2 p_0+k_2^2 p_0+k_3^2 p_0\right)^2 \nonumber \\
	&+
	\left(-k_1 p_1 (k_0+p_0)-k_2 p_2 (k_0+p_0)+(k_3-p_3) (k_3 p_0-k_0 p_3)+k_0 \left(p_1^2+p_2^2\right)+k_1^2 p_0+k_2^2 p_0\right)^2 \nonumber \\
	&+
	\left(-k_0 \left(p_1^2+p_2^2+p_3^2\right)+k_1 p_0 p_1+p_0 (k_2 p_2+k_3 p_3)\right)^2,
\end{align}
\begin{align}\label{O21}
	O_{21}(k)
	=
	0,
\end{align}
\begin{align}
	E_{22}(k)
	&=
	\frac{1}{8} \Big[2 k_0^2 \left(k_1^2-k_1 p_1+k_2^2-k_2 p_2+k_3^2-k_3 p_3-p_0^2\right) \nonumber \\
	&+
	k_0 p_0 \left(-2 k_1^2+4 k_1 p_1-2 k_2^2+4 k_2 p_2-2 k_3^2+4 k_3 p_3+2 p_0^2-p_1^2-p_2^2-p_3^2\right) \nonumber \\
	&+
	k_1^2 \left(3 p_0^2-3 p_1^2-p_2^2-p_3^2\right)+k_1 p_1 \left(-4 k_2 p_2-4 k_3 p_3-4 p_0^2+3 p_1^2+3 p_2^2+3 p_3^2\right) \nonumber \\
	&-
	p_3^2 \left(k_2^2-3 k_2 p_2+3 k_3^2+2 \left(-p_0^2+p_1^2+p_2^2\right)\right)+3 k_2^2 p_0^2-k_2^2 p_1^2-3 k_2^2 p_2^2 \nonumber \\
	&+
	k_3 p_3 \left(-4 k_2 p_2-4 p_0^2+3 p_1^2+3 p_2^2\right)-4 k_2 p_0^2 p_2+3 k_2 p_1^2 p_2+3 k_2 p_2^3+3 k_3^2 p_0^2-k_3^2 p_1^2 \nonumber \\
	&-
	k_3^2 p_2^2+3 k_3 p_3^3+2 p_0^2 p_1^2+2 p_0^2 p_2^2-p_1^4-2 p_1^2 p_2^2-p_2^4-p_3^4\Big],
\end{align}
\begin{align}
	O_{22}(k)
	&=
	\frac{1}{8} \Big[4 k_0^3 p_0+2 k_0^2 \left(-2 k_1 p_1-2 k_2 p_2-2 k_3 p_3-3 p_0^2+p_1^2+p_2^2+p_3^2\right) \nonumber \\
	&+
	k_0 p_0 \left(2 k_1^2+2 k_1 p_1+2 k_2^2+2 k_2 p_2+2 \left(k_3^2+p_0^2\right)+2 k_3 p_3+p_1^2+p_2^2+p_3^2\right)+2 k_1^3 p_1 \nonumber \\
	&-k_1^2 \left(-2 k_2 p_2-2 k_3 p_3+p_0^2+3 p_1^2+p_2^2+p_3^2\right) \nonumber \\
    &-
    k_1 p_1 \left(-2 k_2^2+4 k_2 p_2-2 k_3^2+4 k_3 p_3+2 p_0^2+p_1^2+p_2^2+p_3^2\right) \nonumber \\
	&+
	2 k_2^3 p_2+k_3 p_3 \left(2 k_2^2-4 k_2 p_2+2 k_3^2-2 p_0^2-p_1^2-p_2^2\right)-p_3^2 \left(k_2^2+k_2 p_2+3 k_3^2-2 \left(p_1^2+p_2^2\right)\right) \nonumber \\
	&-
	k_2^2 p_0^2-k_2^2 p_1^2-3 k_2^2 p_2^2+2 k_2 k_3^2 p_2-2 k_2 p_0^2 p_2-k_2 p_1^2 p_2-k_2 p_2^3 \nonumber \\
	&-
	k_3^2 p_0^2-k_3^2 p_1^2-k_3^2 p_2^2-k_3 p_3^3+p_1^4+2 p_1^2 p_2^2+p_2^4+p_3^4\Big].
\end{align}
}


\end{document}